\documentclass[english,10pt, conference, compsocconf]{IEEEtran}
\usepackage[T1]{fontenc}
\usepackage[latin9]{inputenc}
\usepackage{graphicx}

\makeatletter

\IEEEoverridecommandlockouts

\usepackage{subfigure} 
\usepackage{eso-pic} 

\makeatother

\usepackage{babel}
\begin{document}

\title{Social-aware Forwarding in Opportunistic Wireless Networks: Content
Awareness or Obliviousness?}

\author{\IEEEauthorblockN{Waldir Moreira, and Paulo Mendes} \IEEEauthorblockA{SITI,
University Lusófona, Lisbon, Portugal\\
 \{waldir.junior, paulo.mendes\}@ulusofona.pt}\thanks{''This is
the author's preprint version. Personal use of this material is permitted.
However, permission to reprint/republish this material for advertising
or promotion or for creating new collective works for resale or for
redistribution to thirds must be obtained from the copyright owner.
The camera-ready version of this work has been published at AOC 2014,
date of June 2014 and is property of IEEE.''} }
\maketitle
\begin{abstract}
Information-Centric Networking (ICN) has been gaining increasing attention
from the research community as it is able to improve content dissemination
by releasing the dependency on content location. With the current
host-based Internet architecture, networking faces limitations in
dynamic scenarios, due mostly to host mobility. The ICN paradigm mitigates
such problems by releasing the need to have an end-to-end transport
session established during the life time of the data transfer. Moreover,
the ICN concept solves the mismatch between the Internet architecture
and the way users would like to use it: currently a user needs to
know the topological location of the hosts involved in the communication
when he/she just wants to get the data, independently of its location.
Most of the research efforts aim to come up with a stable ICN architecture
in fixed networks, with few examples in ad-hoc and vehicular networks.
However, the Internet is becoming more pervasive with powerful personal
mobile devices that allow users to form dynamic networks in which
content may be exchanged at all times and with low cost. Such pervasive
wireless networks suffer with different levels of disruption given
user mobility, physical obstacles, lack of cooperation, intermittent
connectivity, among others. This paper discusses the combination of
content knowledge (e.g., type and interested parties) and social awareness
within opportunistic networking as to drive the deployment of ICN
solutions in disruptive networking scenarios. With this goal in mind,
we go over few examples of social-aware content-based opportunistic
networking proposals that consider social awareness to allow content
dissemination independently of the level of network disruption. To
show how much content knowledge can improve social-based solutions,
we illustrate by means of simulation some content-oblivious/oriented
proposals in scenarios based on synthetic mobility patterns and real
human traces.\end{abstract}


\section{Introduction}

Information-Centric Networking (ICN) as its own name suggests is driven
by information. That is, data traverses the network according to the
match between its name and the interests that users have in such content,
independently of its location, resulting in an efficient, scalable,
and robust content delivery.

Given its potential, ICN has become of great interest to the research
community \cite{surveyICN}. Currently, there are different approaches
for defining an ICN architecture, such as Data-Oriented Network Architecture
(DONA), Named Data Networking (NDN), and Network of Information (NetInf).
Each of these proposals has its own particularities (e.g., employ
their own naming scheme) and look into different ICN aspects (e.g.,
naming, security, routing, ...) emphasizing few of these aspects according
to the application to which they are being devised.

Yet, these different ICN flavors have some principles in common \cite{surveyICNtrees}:
i) the publish/subscribe paradigm, where users announce the content
that they want to disseminate, and register to receive content made
available by others; ii) the notion of universal caching, upon which
any network node with persistent storage can keep copies of data that
passed through them, or that they requested from other nodes to answer
future requests; iii) security models, which are totally shaped around
the content itself. All these architecture proposals aim at improving
content dissemination in fixed networks (i.e., Internet at large)
with few examples related to ad-hoc and vehicular networking \cite{adhocNDN,CCNmanets,vehicularICN}.

However, wireless devices have become more portable and with increased
capabilities (e.g., processing, storage), which is creating the foundations
for the deployment of pervasive wireless networks, and encompassing
personal devices (e.g. smartphones and tablets). Additionally, wireless
technology has been extended to allow direct communication: vehicle-to-vehicle
- for safety information exchange; device-to-device - aiming at 3G
offloading; Wi-Fi direct - overcome the need for infrastructure entities
(i.e., access points). In such a dynamic scenario, users are prosumers
(i.e., producers and consumers) of information with a high demand
to share/retrieve content anytime and anywhere, independently of the
intermittency level of connectivity, their dynamic behavior, physical
obstacles between them, among others. 

In these dynamic networks, opportunistic contacts among mobile devices
may improve content dissemination, mitigating the effects of network
disruption. This gave rise to the investigation of Opportunistic Networks
(OppNets), of which Delay-Tolerant Networks are an example, encompassing
different forwarding proposals to quickly send data from one point
to another even in the absence of an end-to-end path between them.
Such proposals range from flooding content in the network up to solutions
that take into account the social interactions among users. The latest
set of research findings show that by exploiting social network structures,
social-aware opportunistic networking can indeed improve data forwarding
between two intermittently connect hosts with less cost and latency
than solutions based only on the users' mobility patterns. This is
due to social structures being more stable than connectivity links
created based on the number and frequency of wireless contacts due
to mobility.

By looking at the nature and properties of OppNets and ICN, one can
see the potential of applying content knowledge for driving networking
in OppNets: ICN abstracts the need for establishing an end-to-end
communication session in an environment where end-to-end paths have
little probability of being available. Thus, with ICN, content could
reach the interested parties as they subscribe their interests in
such type of content, releasing the assumption about the existence
of an end-to-end path between any pair of nodes. 

Thus, in this paper we show the advantages of combining social-aware
opportunistic networking with the ICN paradigm, more specifically
the knowledge about the content type and users' interests in such
content. Since we refer ICN and OppNets principles, it is important
to note that the words \emph{information}, \emph{data}, \emph{messages},
and \emph{content}, as well as \emph{users} and \emph{nodes} are used
interchangeably throughout this paper.

The remainder of the paper is structured as follows. Section \ref{sec:SaCbOppRouting}
goes over the social-aware and/or content-based proposals. In Section
\ref{sec:Experiments} we present our experiments that support our
claims, and Section \ref{sec:Conclusions} concludes our work.

\section{Social-aware Forwarding in Opportunistic Wireless Networks\label{sec:SaCbOppRouting}}

Different ICN architecture proposals have emerged considering the
availability, security, and location-independence features of the
new paradigm identified in the Internet such as DONA, NDN, and NetInf.
It is not our intent to list all the efforts to devise an ICN architecture
nor to analyze them as this was already done \cite{surveyICNtrees,surveyICN}.
Yet, our goal is to show that, besides most of the current ICN approaches
targeting communication over fixed networks, the ICN paradigm has
great potential in more dynamic networking scenarios.

As ICN suitably copes with mobility of hosts (and consequently disruptive/intermittent
communications), initial works have investigated its potential when
applied to ad-hoc \cite{adhocNDN,CCNmanets} and vehicular \cite{vehicularICN}
networks. However, these dynamic networks still assume the existence
of at least one end-to-end path between any pair of nodes. In this
case, the challenge posed by such networks is to find the needed end-to-end
path as fast as possible. 

Efforts to devise ICN solutions should target more dynamic scenarios,
in which we cannot assume the existence of end-to-end paths. The ICN
principle has the potential to operate in such disruptive networking
scenarios, since the concept of in-network caching is perfectly aligned
with the concept of persistent storage present in OppNets. Hence,
it would be possible to apply content knowledge (e.g., type and interested
parties) in OppNets by revisiting the breadcrumb approach (data is
forwarded over an interface where previously an interest was received)
to operate based on contact opportunities and not on stable network
interfaces. 

However, if we look at available opportunistic networking solutions,
all of them aim to forward data from point A to point B considering
single-copy forwarding or content replication at different levels
based on node encounter, resource usage, or social similarity. That
is, opportunistic networking operates based on the identification
of the hosts (source, destination) and is not based on the transported
content, despite the fact that it has been shown how dynamic scenarios
can benefit (performance improvements and wise resource usage) from
considering the content properties while performing forwarding \cite{ICNoppNet2008A,ICNoppNet2008B}.

From the opportunistic networking solutions, the social-aware family
of replication-based proposals gained attention given its ability
to avoid the volatile property of mobility. That is, instead of considering
the number and frequency of contacts due to the mobility of hosts,
such approaches take into account more stable social aspects (e.g.,
common social groups and communities, node popularity, levels of centrality,
shared interests, and future social interactions), aiming to reduce
the cost of opportunistic forwarding. 

By looking at this social-aware family, one can distinguish between
proposals that are completely unaware of content information (i.e.,
content-oblivious) and those that consider different levels of content
knowledge (i.e., content-oriented) while taking forwarding decisions.
For the sake of simplicity, we only list the most relevant and latest
proposals.

Among the social-aware content-oblivious proposals, we analyzed \textit{Bubble
Rap} \cite{bubble2011}, \emph{CiPRO} \cite{cipro}, and \emph{dLife}
\cite{dlife}. 

\textit{Bubble Rap} combines the node centrality with the notion of
community to make forwarding decisions. The centrality metric identifies
hub nodes inside (i.e.\,local) or outside (i.e., global) communities.
Messages are replicated based on global centrality until they reach
the community of the destination host (i.e., a node belonging to the
same community). Then, it uses the local centrality to reach the destination
inside the community.

\emph{CiPRO} considers the time and place nodes meet throughout their
routines.\emph{ CiPRO }holds knowledge of nodes (e.g., carrier's name,
address, nationality, ...) expressed by means of profiles that are
used to compute the encounter probability among nodes in specific
time periods. Nodes that meet occasionally get a copy of the message
only if they have higher encounter probability towards its destination.
If nodes meet frequently, history of encounters is used to predict
encounter probabilities for efficient broadcasting of control packets
and messages.

\emph{dLife} takes into account the dynamism of users' behavior found
in their daily life routines to aid forwarding. The goal is to keep
track of the different levels of social interactions (in terms of
contact duration) nodes have throughout their daily activities in
order to infer how well socially connected they are in different periods
of the day. 

Regarding social-aware content-oriented proposals, we analyzed \textit{SocialCast}
\cite{socialcast},\emph{ ContentPlace} \cite{contentplace}, and
\emph{SCORP} \cite{scorp}. \textit{SocialCast} considers the interest
shared among nodes. It devises a utility function that captures the
future co-location of the node (with others sharing the same interest)
and the change in its connectivity degree. Thus, the utility function
measures how good message carrier a node can be regarding a given
interest. \textit{SocialCast} functions are based on the publish-subscribe
paradigm, where users broadcast their interests, and content is disseminated
to interested parties and/or to new carriers with high utility.

\emph{ContentPlace} considers information about the users' social
relationships to improve content availability. It computes a utility
function for each data object considering: i) the access probability
to each object and the involved cost in accessing it; ii) the social
strength of the user towards the different communities which he/she
belongs to and/or has interacted with. The idea is having the users
to fetch data objects that maximize the utility function with respect
to local cache limitations, and choosing those objects that are of
interest to users and can be further disseminated in the communities
they have strong social ties.

\emph{SCORP} considers the type of content and the social relationship
between the parties interested in such content type. \emph{SCORP}
nodes are expected to receive and store messages considering their
own interests as well as interests of other nodes with whom they have
interacted before. Data forwarding takes place by considering the
social weight of the encountered node towards nodes interested in
the message that is about to be replicated.

Generally speaking, although OppNets can provide communication support
when facing disruptive networks, ICN has the potential to cope with
disruptive/intermittent communications since it does not require the
establishment of associations between the source and destination of
content. Both paradigms mainly differ in what concerns forwarding:
ICN forwarding considers data names and OppNets forwarding focuses
on hosts. Thus, by combining content knowledge with social-aware forwarding
may increase the performance of data exchange in OppNets, since: i)
nodes sharing interests have higher probability to meet each other;
and ii) social-awareness results in fast dissemination given the contact
opportunities among nodes.

\section{Content Awareness or Obliviousness: Which way to go?\label{sec:Experiments}}

Based on experiments, we analyze what is the impact of combining content
and social awareness to forward data in opportunistic networks. We
consider two social-aware content-oblivious opportunistic routing
solutions (i.e., \emph{dLife} and \emph{Bubble Rap}) and one social-aware
content-oriented opportunistic routing solution (i.e.,\emph{ SCORP}).
It is worth mentioning that our goal is not to show which proposal
is the best; instead, we want to show the gains of combining content
knowledge and social awareness, and for that we chose the benchmark
proposals that are readily available for the Opportunistic Network
Environment (ONE) simulator \cite{one}.

Two scenarios are considered: human traces to observe the impact of
network load; and synthetic mobility to study how proposals cope with
different mobility levels, from high mobile to near static nodes.
This section starts by presenting the evaluation methodology and experiment
settings, then followed by the results in the considered scenarios.

\subsection{Methodology and Experimental Settings\label{sub:Evaluation-Methodology}}

Results are presented with a 95\% confidence interval and in terms
of averaged delivery probability (i.e., ratio between the number of
delivered messages and the number of messages that should have been
delivered), cost (i.e., number of replicas per delivered message),
and latency (i.e., time elapsed between message creation and delivery). 

The used CRAWDAD traces \cite{cambridge-haggle-imote-content-2006-09-15}
corresponds to contacts of 36 students during their daily activities.

The synthetic mobility scenario simulates a 4-day interaction between
3 groups ($A$, $M$, and $B$) of 50 people each, who carry nodes
equipped with 250-Kbps Bluetooth interfaces, and follow the \emph{Shortest
Path Map Based Movement} model (i.e., nodes choose destinations and
reach them by using the shortest path) with speed up to 1.4 m/s. By
varying the node pause times between 100 and 100000 seconds, we have
different levels of mobility (varying from 3456 to 3.4 movements in
the simulation). 

Proposals experience the same load and number of messages that must
reach the destinations. In the trace scenario, the \emph{Bubble Rap}/\emph{dLife}
source sends 1, 5, 10, 20 and 35 different messages to each of the
35 destinations, while the \emph{SCORP} source creates 35 messages
with unique content types, and the receivers are configured with 1,
5, 10, 20, and 35 randomly assigned interests. Thus, we have a total
of 35, 175, 350, 700, and 1225 generated messages. The \emph{msg/int}
notation represents the number of messages sent by \emph{Bubble Rap}
and \emph{dLife} sources, or the number of interests of each of the
\emph{SCORP} receivers.

In the synthetic mobility scenario, 200 messages are generated. With
\emph{Bubble} \emph{Rap }and \emph{dLife}, node $0$ (group \emph{$A$})
generates 100 messages to nodes in groups $B$ and $M$, and node
$100$ (group $B$) generates 100 messages to nodes in groups $A$
and $M$. For \emph{SCORP}, each group has different interests: group
$A$ (\emph{reading}), group $B$ (\emph{games}), and group $M$ (\emph{reading}
and \emph{games}). The source nodes, $0$ and $100$, generate only
one message for each content type, \emph{game} and \emph{reading}.
This guarantees the same number of messages expected to be received,
i.e., 200. 

Concerning TTL, data-centric networking is expected to allow content
to reach interested nodes independently of how long it takes, due
to the assumption about persistent storage. So, in these experiments,
TTL is set to avoid messages being discarded due to expiration (i.e.,
the length of the experiment: 3 weeks and 4 days for for trace-based
and synthetic mobility scenarios, respectively). Message size ranges
from 1 to 100 kB. Despite nodes may have plenty of storage, we consider
nodes having different capabilities (i.e., smartphones). Thus, nodes
have buffers limited to 2 MB as we consider that nodes may not be
willing to share all their storage space. These settings follow the
specification of the Universal Evaluation Framework \cite{latincom}
to guarantee fairness throughout the evaluation process. 

Since \emph{Bubble Rap}/\emph{dLife} sources generate more messages,
in the trace-based scenario node $0$ has no buffer restriction and
message generation varies with the load: 35 messages/day rate (load
of 1, 5, and 10 messages), and 70 and 140 messages/day rates (load
of 20 and 35 messages, respectively). In the synthetic mobility scenario,
all source nodes have restricted buffer, but rate is of 25 messages
every 12 hours. This is done so that \emph{Bubble Rap}/\emph{dLife}
do not discard messages prior to even trying exchange/deliver them
given the buffer constraint.

As for proposals, \emph{Bubble Rap }uses the K-Clique and cumulative
window algorithms for community formation and centrality computation
as in \cite{bubble2011}. As for \emph{dLife }and \emph{SCORP}, both
consider 24 daily samples (i.e., each of one hour) as mentioned in
\cite{scorp}.

\subsection{Impact of Network Load\label{sub:Impact-of-Network}}

\begin{figure}[h]
\centering{}\includegraphics[scale=0.7]{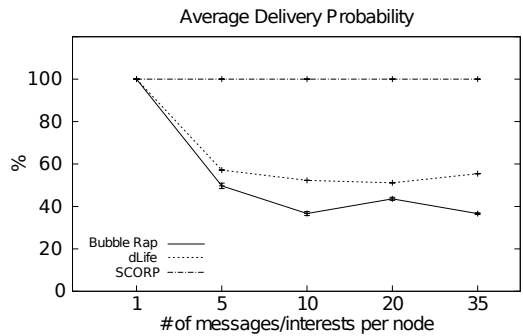}\protect\caption{\label{fig:2}Delivery under different network loads}
\end{figure}

This section presents the impact that different levels of network
load have on the performance of\emph{ }the content-oblivious and -oriented
forwarding proposals. Fig. \ref{fig:2} presents the average delivery
probability with different messages/interests being generated.

In the 1 msg/int configuration, formed communities comprise almost
all nodes. This means that each node has high probability to meet
any other node, which is advantageous for \emph{Bubble Rap} since
most of its deliveries happen to nodes sharing communities. Due to
the dense properties of the network, \emph{dLife} and \emph{SCORP
}take advantage of direct delivery: 57\% and 51\% of messages, respectively,
are delivered directly to destinations.

As load increases, \emph{Bubble Rap }has an 50\% decrease in delivery
performance. This occurs since it relies on communities to perform
forwardings, and consequently buffer space becomes an issue. To support
this claim, we estimate buffer usage for the 5 msg/int configuration:
there is an average of 80340.7 forwardings, and if this number is
divided by the number of days (12%
\footnote{~In simulation it is worth \textasciitilde{}12 days of communications.%
}) and by the number of nodes (35, source not included), we get an
average of 191.28 replications per node. Multiplied by the average
message size (52kB), the buffer occupancy is roughly 9.94 MB in each
node, which exceeds\textbf{ }the 2MB allowed (cf. Sec. 4.1). This
estimation is for a worst case scenario where \emph{Bubble Rap} spreads
copies to every encountered node. Since this cannot happen, as \emph{Bubble
Rap} also relies on local centrality to reduce replication, buffer
exhaustion is really an issue given that messages are replicated to
fewer nodes and not to all as in our estimation. As more messages
are generated, replication increases: this causes the spread of messages
that potentially take over forwarding opportunities from other messages,
reducing \emph{Bubble Rap}'s delivery capability.

\emph{dLife} has a 43\% performance decrease when network load increases,
as it takes time to have an accurate view of the social weights. This
leads to forwardings that never reach destinations given the contact
sporadicity. For the 10 msg/int configuration, \emph{dLife} also experiences
buffer exhaustion: estimated consumption is 2.17 MB per node. Still,
by considering social weights or node importance allows \emph{dLife}
a more stable behavior than \emph{Bubble Rap}. 

Since content is only replicated to nodes that are interested in it
or that have a strong social interaction with other nodes interested
in such content, the delivery capability of \emph{SCORP} raises as
the ability of nodes to become a good carrier increases (i.e., the
more interests a node has, the better it is to deliver content to
others, since they potentially share interests). The maximum estimated
buffer consumption of \emph{SCORP} is of 0.16 MB (35 msg/int). 

\begin{figure}[h]
\centering{}\includegraphics[scale=0.7]{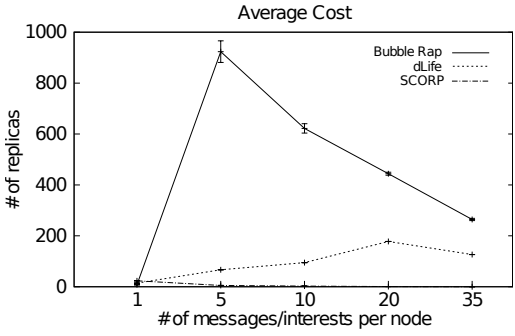}\protect\caption{\label{fig:3}Cost under different network loads}
\end{figure}

Fig. \ref{fig:3} presents the average cost behavior. In the 1 msg/int
configuration, all proposals create very few replicas to perform a
successful delivery, 7.95 (\emph{Bubble Rap}), 14.32 (\emph{dLife}),
and 23.46 (\emph{SCORP}), as they rely mostly on shared communities
and/or direct deliveries. We also observe that \emph{SCORP} produces
more replicas than \emph{dLife}, since \emph{SCORP} nodes with interest
in a specific content of a message not only process it, but also replicate
it to other interested nodes, thus creating extra replicas.

For the 5, 10, 20 and 35 msg/int configurations, replication is directly
proportional to the load. Thus, cost is expected to increase as load
increases, as seen with \emph{Bubble Rap }and \emph{dLife}. Despite
their efforts, these replications do not improve their delivery probabilities,
contributing only to the associated cost for performing successful
deliveries.

The cost peaks relate to the message creation time and contact sporadicity:
when a message is created in a period of high number of contacts,
resulting in much more replications. This is more evident with \emph{Bubble
Rap} as it relies on shared communities to forward: as mentioned earlier,
most of the communities comprise almost all nodes, which increases
its replication rate.

With more interests, a \emph{SCORP} node can serve as a carrier for
a larger number of nodes. Consequently, the observed extra replicas
make the proposal rather efficient: \emph{SCORP} creates an average
of 6.39 replicas across all msg/int configurations, while \emph{Bubble
Rap }and \emph{dLife} produce an average 452.41 and 96 replicas, respectively.

\begin{figure}[h]
\centering{}\includegraphics[scale=0.7]{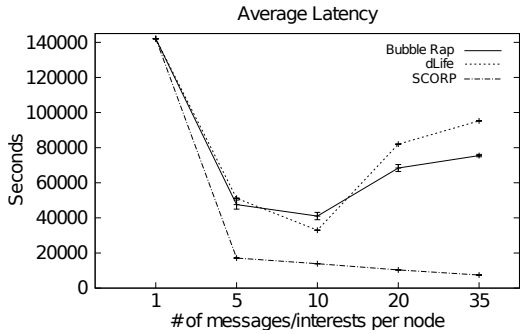}\protect\caption{\label{fig:4}Latency under different network loads}
\end{figure}

Fig. \ref{fig:4} shows the average latency that messages experience.
The latency peak in the 1 msg/int configuration refers to the message
generation time: some messages are created during periods where very
few contacts (and sometimes none) take place followed by long periods
(12 to 23 hours) with almost no contact. Consequently, messages are
stored longer, contributing to the increase of the overall latency.
This effect is mitigated as the load increases with messages being
created almost immediately before a high number of contacts take place. 

Since latency is in function of the delivered messages, the decrease
and variable behavior of \emph{Bubble Rap }and \emph{dLife} is due
to their delivery rates decrease and increase, and also to their choices
of next forwarders that may take longer to deliver content to destinations.
\emph{SCORP} experiences latencies up to approx. 90.2\% and 92.2\%
less than \emph{Bubble Rap} and \emph{dLife}, respectively. The ability
of a node to deliver content increases with the number of its interests.
Thus, a node can receive more messages when it is interested in their
contents, and consequently becomes a better forwarder since the probability
of coming into contact with other nodes sharing similar interests
is very high, thus reducing latency.

\subsection{Impact of Mobility Rates\label{sub:Impact-of-Mobility}}

This section presents the impact that node mobility has on the proposals.
As ICN approaches are devised to rather stationary scenarios (e.g.,
Internet), we observe the performance of the analyzed proposals under
a variety of node mobility, including a near-static scenario.

\begin{figure}[h]
\centering{}\includegraphics[scale=0.7]{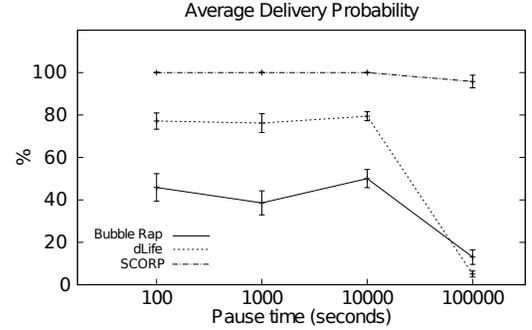}\protect\caption{\label{fig:5}Delivery under varied mobility rates}
\end{figure}

Fig. \ref{fig:5} presents the average delivery probability. Given
the community formation characteristic of this scenario, \emph{Bubble
Rap} relies mostly on the global centrality to deliver content. By
looking at centrality \cite{bubble2011}, we observe very few nodes
(out of the 150) with global centrality that can actually aid in forwarding,
i.e., 19.33\% (29 nodes), 10.67\% (16 nodes), 21.33\% (32 nodes),
and 2\% (3 nodes) for 100, 1000, 10000, and 100000 pause time configurations,
respectively. So, these nodes become hubs and given buffer constraint
and long TTL (i.e., messages created earlier take the opportunity
of newly created ones), message drop is certain, directly impacting
\emph{Bubble Rap}. 

Given the high number of contacts, the computation of social weight
and node importance done by \emph{dLife} takes longer to reflect reality:
thus, \emph{dLife} replicates more and experiences buffer exhaustion.
Indeed, social awareness is advantageous, but still not enough to
reach optimal delivery rate in such conditions.

Independent of the number of contacts among nodes, \emph{SCORP} can
still identify nodes that are better related to others sharing similar
interests, reaching optimal delivery rate for 100, 1000, and 10000
pause time configurations. By considering nodes' interest in content
and their social weights, \emph{SCORP} does not suffer as much with
node mobility as \emph{dLife} and \emph{Bubble} \emph{Rap}.

With 100000 seconds of pause time, the little interaction happening
in a sporadic manner (with intervals between 20 and 26 hours) affects
\emph{Bubble Rap}, \emph{dLife }and \emph{SCORP} as they depend on
such interactions to compute centrality, node importance, and social
weights, as well as to exchange/deliver content.

\begin{figure}[h]
\centering{}\includegraphics[scale=0.7]{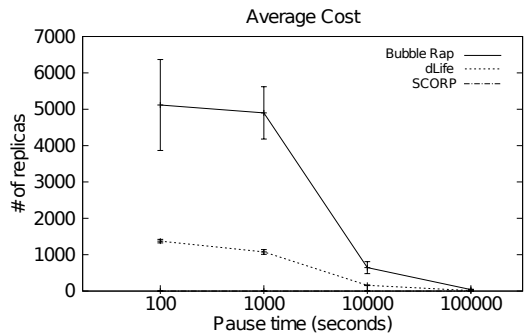}\protect\caption{\label{fig:6}Cost under varied mobility rates}
\end{figure}

Fig. \ref{fig:6} presents the average cost behavior. As pause time
increases, the number of contacts among nodes decreases, providing
all solutions with the opportunity to have a stable view of the network
in terms of their social metrics with 100, 1000, and 10000 seconds
of pause time. This explains the cost reduction experienced by \emph{Bubble
Rap} and \emph{dLife}: both are able to identify the best next forwarders,
which results in the creation of less replicas to perform a successful
delivery. 

\emph{SCORP} has a very low replication rate (average of 0.5 replicas)
given its choice to replicate based on the interest that nodes have
on content and on their social weight towards other nodes interested
in such content. When the intermediate node has an increased number
of interests (i.e., by having different interests, the node can potentially
deliver more content) as observed in Sec. \ref{sub:Impact-of-Network},
replication costs are even lower. Furthermore, \emph{SCORP }suitably
uses buffer space with an estimated average occupancy of 0.03 MB per
node per day.

With 100000 seconds of pause time, as cost is in function of delivered
messages (and deliveries are very low, due to contact sporadicity),
proposals have a low cost. 

\begin{figure}[h]
\centering{}\includegraphics[scale=0.7]{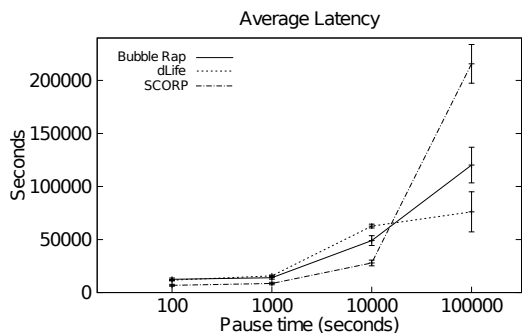}\protect\caption{\label{fig:7}Latency under varied mobility rates}
\end{figure}

As expected (cf. Fig. \ref{fig:7}), latency increases as node mobility
decreases: encounters are less frequent, and so content must be stored
for longer times. Also, the time that the social metrics take to converge
(i.e., a more stable view of the network in terms of centrality, social
weight, and node importance) contributes for the increase in the experienced
latency. The highest increase in latency with 100000 seconds of pause
time is due to contacts happening in a sporadic fashion with intervals
between them of up to 26 hours, thus proposals take much longer to
perform a delivery.

\section{Conclusions \label{sec:Conclusions}}

Information-Centric Networking has become of great interest to the
research community and we have witnessed a number of efforts towards
the definition of what the ICN architecture should be. Yet, most of
the ICN application scenarios encompass fixed networks (i.e., Internet
at large) with few ad-hoc and vehicular networking examples \cite{adhocNDN,CCNmanets,vehicularICN}.

Concerning more dynamic networks, opportunistic networking is being
investigated to support forwarding between any two points even in
the absence of end-to-end paths. Within all opportunistic networking
solutions, one trend for dealing with network disruption is considering
the social interactions among users. Social-aware approaches have
indeed shown great potential considering different types of social
metrics. 

By looking at the nature of OppNets and the properties of ICN, one
can see the potential of applying content knowledge (i.e., type and
interested parties) for driving networking in opportunistic networks:
the ICN paradigm abstracts the need for establishing an end-to-end
communication session in an environment where end-to-end paths have
little probability of being available. Nevertheless, one open question
is related to the impact that content awareness may bring to forwarding
in opportunistic networks. 

Therefore, this paper discusses on the advantages of building social-
and content-aware forwarding schemes for networking in disruptive
scenarios. Our experiments show that by building a content-oriented
social-aware opportunistic forwarding scheme, delivery in disruptive
networks can be improved by 60\% while latency and cost can be reduced
by 75\% and 90\% respectively, when compared to content-oblivious
forwarding schemes.

\bibliographystyle{ieeetr}
\bibliography{bib-or}

\begin{thebibliography}{10}

\bibitem{surveyICN}
B.~Ahlgren, C.~Dannewitz, C.~Imbrenda, D.~Kutscher, and B.~Ohlman, ``A survey
  of information-centric networking,'' {\em Communications Magazine, IEEE},
  vol.~50, no.~7, pp.~26 --36, July, 2012.

\bibitem{surveyICNtrees}
A.~Ghodsi, S.~Shenker, T.~Koponen, A.~Singla, B.~Raghavan, and J.~Wilcox,
  ``Information-centric networking: seeing the forest for the trees,'' in {\em
  Proceedings of HotNets-X}, Cambridge, USA, November, 2011.

\bibitem{adhocNDN}
M.~Meisel, V.~Pappas, and L.~Zhang, ``Ad hoc networking via named data,'' in
  {\em Proceedings of MobiArch}, Chicago, USA, September, 2010.

\bibitem{CCNmanets}
S.~Oh, D.~Lau, and M.~Gerla, ``Content centric networking in tactical and
  emergency manets,'' in {\em Proc. of IFIP Wireless Days}, Venice, Italy,
  October, 2010.

\bibitem{vehicularICN}
M.~Aguilera~Leal, M.~Rockl, B.~Kloiber, F.~de~Ponte~Muller, and T.~Strang,
  ``Information- centric opportunistic data dissemination in vehicular ad hoc
  networks,'' in {\em Proceedings of ITSC}, Madeira Island, Portugal,
  September, 2010.

\bibitem{ICNoppNet2008A}
C.~Boldrini, M.~Conti, and A.~Passarella, ``Context and resource awareness in
  opportunistic network data dissemination,'' in {\em Proceedings of WoWMoM},
  Newport Beach, USA, June, 2008.

\bibitem{ICNoppNet2008B}
C.~Boldrini, M.~Conti, and A.~Passarella, ``Modelling data dissemination in
  opportunistic networks,'' in {\em Proceedings of CHANTS}, San Francisco, USA,
  September, 2008.

\bibitem{bubble2011}
P.~Hui, J.~Crowcroft, and E.~Yoneki, ``Bubble rap: Social-based forwarding in
  delay-tolerant networks,'' {\em IEEE Transactions on Mobile Computing},
  vol.~10, no.~11, pp.~1576--1589, November, 2011.

\bibitem{cipro}
H.~A. Nguyen and S.~Giordano, ``Context information prediction for social-based
  routing in opportunistic networks,'' {\em Ad Hoc Netw.}, vol.~10, no.~8,
  pp.~1557--1569, November, 2012.

\bibitem{dlife}
W.~Moreira, P.~Mendes, and S.~Sargento, ``Opportunistic routing based on daily
  routines,'' in {\em Proceedings of WoWMoM}, San Francisco, USA, June, 2012.

\bibitem{socialcast}
P.~Costa, C.~Mascolo, M.~Musolesi, and G.~P. Picco, ``Socially-aware routing
  for publish- subscribe in delay-tolerant mobile ad hoc networks,'' {\em IEEE
  J.Sel. A. Commun.}, vol.~26, no.~5, pp.~748--760, June, 2008.

\bibitem{contentplace}
C.~Boldrini, M.~Conti, and A.~Passarella, ``Design and performance evaluation
  of contentplace, a social-aware data dissemination system for opportunistic
  networks,'' {\em Comput. Netw.}, vol.~54, no.~4, pp.~589--604, March, 2010.

\bibitem{scorp}
W.~Moreira, P.~Mendes, and S.~Sargento, ``Social-aware opportunistic routing
  protocol based on user's interactions and interests,'' in {\em Proceedings of
  AdHocNets}, Barcelona, Spain, October, 2013.

\bibitem{one}
A.~Ker\"{a}nen, J.~Ott, and T.~K\"{a}rkk\"{a}inen, ``The one simulator for dtn
  protocol evaluation,'' in {\em Proceedings of SIMUTools}, Rome, Italy, March,
  2009.

\bibitem{cambridge-haggle-imote-content-2006-09-15}
J.~Scott, R.~Gass, J.~Crowcroft, P.~Hui, C.~Diot, and A.~Chaintreau,
  ``{CRAWDAD} trace cambridge /haggle/imote/content (v. 2006-09-15).''
  Available at http://crawdad.cs.dartmouth.edu/, 2006.

\bibitem{latincom}
W.~Moreira, P.~Mendes, and S.~Sargento, ``Assessment model for opportunistic
  routing,'' in {\em Proceedings of LATINCOM}, Belem, Brazil, October, 2011.

\end{thebibliography}

\end{document}